\documentclass[sigconf]{acmart}
\usepackage{listings}

\AtBeginDocument{%
  }

\acmYear{2026}
 \setcopyright{cc}
\acmConference[CHI '26]{Proceedings of the 2026 CHI Conference on Human Factors in Computing Systems}{April 13--17, 2026}{Barcelona, Spain}
\acmBooktitle{Proceedings of the 2026 CHI Conference on Human Factors in Computing Systems (CHI '26), April 13--17, 2026, Barcelona, Spain}
\acmPrice{}
\acmDOI{10.1145/3772318.3790310}
\acmISBN{979-8-4007-2278-3/2026/04}

\begin{document}

\title{\textit{I, Robot?} Exploring Ultra-Personalized AI-Powered AAC; an Autoethnographic Account}

\author{Tobias Weinberg}
\affiliation{
  \institution{Cornell Tech}
  \city{New York}
  \country{USA}}
\email{tmw88@cornell.edu}

\author{Ricardo E. Gonzalez Penuela}
\affiliation{%
  \institution{Cornell Tech}
  \city{New York}
  \country{USA}}
\email{reg258@cornell.edu}

\author{Stephanie Valencia}
\authornote{Both authors contributed equally to the paper.}
\affiliation{%
  \institution{University of Maryland}
  \city{College Park, MD}
  \country{USA}}
\email{sval@umd.edu}

\author{Thijs Roumen}
\authornotemark[1]
\affiliation{%
  \institution{Cornell Tech}
  \city{New York}
  \country{USA}}
\email{thijs.roumen@cornell.edu}

\renewcommand{\shortauthors}{Weinberg et al.}

\begin{abstract}
 Generic AI auto-complete for message composition often fails to capture the nuance of personal identity, requiring editing. While harmless in low-stakes settings, for users of Augmentative and Alternative Communication (AAC) devices, who rely on such systems to communicate, this burden is severe. Intuitively, the need for edits would be lower if language models were personalized to the specific user's communication. 
 
While personalization is technically feasible, it raises questions about how such systems affect AAC users’ agency, identity, and privacy. We conducted an autoethnographic study in three phases: (1) seven months of collecting all the lead author’s AAC communication data, (2) fine-tuning a model on this dataset, and (3) three months of daily use of personalized AI suggestions. We observed that: logging everyday conversations reshaped the author’s sense of agency, model training selectively amplified or muted aspects of his identity, and suggestions occasionally resurfaced private details outside their original context. 
 
Our findings show that ultra-personalized AAC reshapes communication by continually renegotiating agency, identity, and privacy between user and model. We highlight design directions for building personalized AAC technology that supports expressive, authentic communication.

\end{abstract}




\begin{teaserfigure}
\centering
  \includegraphics[width=0.63\textwidth, alt={}]{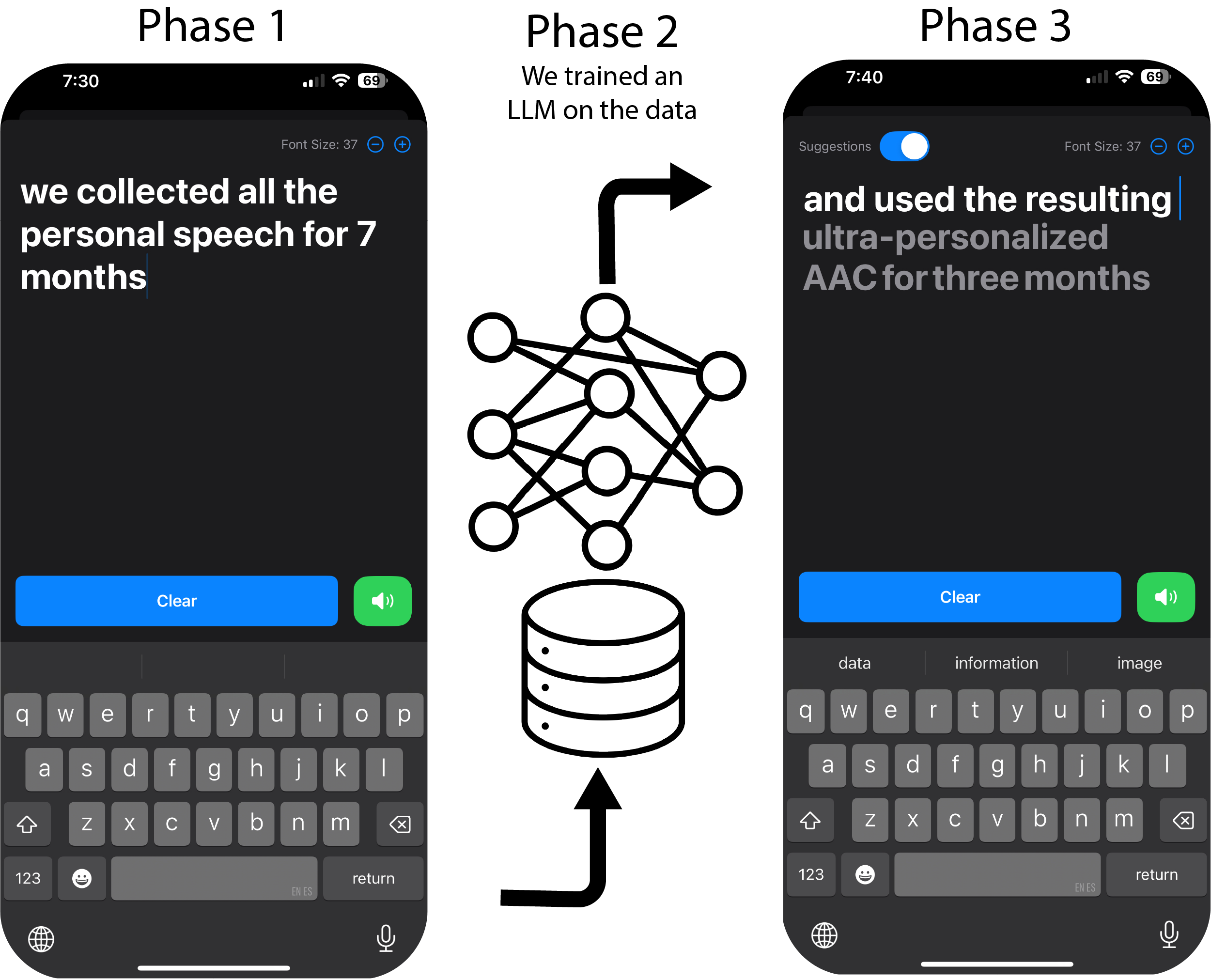}
  \caption{We explored the socio-technical implications of ultra-personalization in AAC across three phases.}
  \Description{Three-panel study overview. Phase 1: app screen showing the text 'collecting the author’s AAC speech for 7 months'. Phase 2: database and neural-network icons with arrows indicate training an LLM on the collected data. Phase 3: app screen showing the text 'using the resulting' and the trained completed in in-line ghost text 'ultra-personalized AAC for 3 months'.}
  \label{fig:teaser}
\end{teaserfigure}

% \makeatletter
% \def\@ACM@copyright@check@cc{}
% \makeatother
\maketitle

\section{Introduction}

For people with speech impairments who type to speak, composing messages with just a few clicks is not a luxury but a necessity~\cite{waller_telling_2019, seale2020interaction}. Yet, generic auto-complete systems for composing messages often fall short: they rarely capture the nuance of personal identity, forcing users to spend significant effort editing suggestions~\cite{valencia_less_2023, valencia2024compa}. For users of Augmentative and Alternative Communication (AAC) devices, who rely on such systems for their everyday speech, this becomes a heavy burden~\cite{kane_at_2017, waller2025think}. Intuitively, the need for text editing would be significantly lower if the model was trained on the specific user's communication as opposed to a generic model. While personalization using large language models has been shown to be technically feasible~\cite{salemi2024lamp, chen2024large, cai_speakfaster_2023}, 
it is essential to understand how such systems may reshape AAC users’ everyday communication before deploying them in real-world settings~\cite{dash2025gets, shneiderman2020human, reisman2018algorithmic}.

Within the field of AAC research, several AI-based techniques have been proposed to predict or expand on users' typed messages~\cite{valencia2024compa, valencia_less_2023}. \textit{Kwickchat}~\cite{shen_kwickchat_2022} a sentence-based text entry system that allows AAC users to generate replies with minimal input by predicting sentences based on keyword entry. \citet{cai_using_2023} demonstrated the potential of LLMs to abbreviate input for eye-gaze, and further showed that incorporating conversational context significantly improved the efficiency~\cite{cai2022context}. \citet{weinberg2025why} found that in time-pressured scenarios, some AAC users were willing to trade a degree of authorship to the LLM for speed gain, while others expressed concerns about authenticity and the risk of AI altering their self-perception. Other recent community consultation also highlighted concerns about privacy, authorship authenticity, and data governance~\cite{griffiths2024use, valencia_less_2023}, warning that LLM-enabled AAC risk reducing opportunities for co-construction if adopted uncritically~\cite{griffiths2025ai}.

Outside of AAC, personalization of LLMs has been shown to enhance relevance and fluency of AI-assisted writing, but also raises concerns around authorship, tone, and control~\cite{peng2024reviewllm, liu2024llms+, salemi2024lamp, yeh2024ghostwriter, li2023teach, kadoma2024role}. Furthermore, \citet{kane_at_2017} emphasizes that AAC users often go to great lengths to preserve self-expression, using humor, timing, and personalization to construct a sense of identity.  
 Building on these insights, personalized LLMs for AAC systems promise faster, more fluent communication, yet they also introduce tensions: AI-suggestions can diminish a user’s sense of agency, amplifying certain identity cues can risk misrepresentation, and breaching privacy by exposing sensitive information out of context. Studying AI personalization use in practice is key to help us identify benefits and risks for AAC users.

First-person accounts have been shown to have the potential to provide a perspective grounded in the current capabilities and flaws of AI in the lives of the users~\cite{glazko2023autoethnographic, glazko2025autoethnographic}. In this work, we empirically examine how an ultra-personalized, AI-powered AAC system (defined as a user-specific LLM with an interface and interaction design built around that user’s unique communication style) shapes everyday communication when trained on a single user’s full AAC history and evaluated through an in-depth autoethnographic account. We analyze how such a system influences the user’s agency (control over one’s words), identity (representation of cultural, linguistic, and tonal patterns), and privacy (what personal information is logged, learned, and resurfaced across different social settings). We explored the space by asking the following questions: How does continuous data logging and interacting with the AI shape the user’s agency over their own speech? In what ways does the personalized model reflect the user’s identity? And how do personalized suggestions risk violating privacy when they resurface in the wrong moments?

We approach these questions in three phases: (1)~the lead author (who has a speech impairment and is a full-time AAC user) collected seven months of all his personal speech data, (2)~we filtered the data, trained a language model, and built a simple AAC app implementing the trained model (3)~finally, the lead author used that app as his main AAC device for 3 months collecting and usage data. The lead author reflected on each phase through diary logs, providing a grounded perspective in his experience across the process.

Our reflections reveal dynamics around how personalization shapes agency, identity, and privacy. 
In Phase 1, we found that collecting the speech data led the author to avoid certain expressions like swear words or certain jokes. Such self-censorship diminished the authenticity, removing slang and dark humor from the dataset, constraining the author's agency to express freely. Curating the dataset in Phase 2 removed expressions that the author was not comfortable delegating to the model to reproduce on its own, which effectively produced a socially “well-behaved" model that reflected only the "good" part of the author's identity. Furthermore, in the deployment and evaluation (Phase 3), while the model successfully mimicked the author's multicultural identity in daily conversation, the author experienced privacy breaches when the model occasionally resurfaced intimate biographical details and cultural markers in inappropriate contexts.

These findings highlight both the potential and the risks of ultra-personalized AI in AAC: It requires careful attention to how such systems respond in real-time to evolving conversations (responsiveness), whether their suggestions align with the norms and expectations of a given interaction (social context), and how much control users retain over what is said and how it is said (user agency). Without this alignment, even highly personalized outputs may feel intrusive, inauthentic, or might misrepresent the user, undermining the very identity they aim to support.%}{such systems must continually balance the user’s agency, identity, and privacy. Without this balance, even highly accurate suggestions can feel intrusive, inauthentic, or misaligned, ultimately undermining the very identity they are designed to support.}

\bigskip
\bigskip

\section{Contributions, Benefits, and Limitations}

This paper makes the following contributions:

\begin{itemize}
\item \textbf{Empirical insight into ultra-personalized AAC:} Through an autoethnographic study grounded in a multi-month deployment, we provide a nuanced account of how a large language model (LLM) trained on a single AAC user’s communication data shaped their agency, identity, and privacy boundaries.

\item \textbf{Identification of tensions in }personalized AI. We surface trade-offs to ultra-personalized systems, showing how personalization can constrain \emph{agency} through self-censorship, reshape \emph{identity} by erasing or amplifying certain aspects of the user’s voice, and threaten \emph{privacy} when models overstep contextual boundaries or surface culturally specific content at inappropriate moments.

\item \textbf{Positioning lived experience in AAC research:} By documenting the evolving relationship between user and model over three months of daily use of personalized AI suggestions, we highlight real-world tensions in personalization, authorship, and data use that may not emerge in lab-based or short-term studies. We hope this work inspires further use of experiential, longitudinal methods to surface the complexities of AI adoption in assistive technologies, before deployment at scale.
\end{itemize}

This study contributes early, experience-driven insight into how ultra-personalized AI might enhance expressivity, efficiency, and cultural relevance for AAC users. By examining how personalization reshapes agency, identity, and privacy over time, we offer design implications for systems that aim to preserve user intent while accelerating message composition.

\subsection{Limitations}

This work is grounded in the lived experience of a single AAC user-researcher who only experiences speech production and minor motor disabilities and types using a full (digital) keyboard. We do not claim generalizability across the wider population of AAC users. Proposed conclusions assume literate AAC users without cognitive impairments who communicate primarily through text-entry–based speech-generating devices. It assumes the ability to compose, review, and selectively accept suggestions, making it most applicable to conversational contexts that support visual verification and user-driven message editing.

The autoethnographic focus of this study allowed us to surface fine-grained dynamics that would be difficult to observe otherwise. It also has important limitations. Many of the phenomena we describe, such as self-censorship or shifts in authorship, are difficult to access through external observation. At the same time, the account necessarily reflects a single trajectory, shaped by the author’s individual communication patterns, cultural background, and research motivations. Future work should explore multi-participant longitudinal studies to generalize these insights to broader populations of AAC users.

We also note that the technical implementation of our prototype, such as model architecture, training method, or system performance, is not a core contribution of this work. We took a pragmatic approach to building a functioning personalized system for long-term use, rather than optimizing for state-of-the-art AI performance. Environmental factors (e.g., interface lag or network issues) occasionally limited the system’s responsiveness, and future offline-capable or more robust implementations may behave differently. Our conclusions are therefore not tied to the specific accuracy of the model, but instead focus on the broader experiential implications of living with a personalized AI over time.

Our deployment relied on a cloud-hosted LLM, and although we observed occasional delays in low-connectivity environments, we did not systematically measure latency or quantify its impact. Similarly, while our interaction logs capture acceptance and rejection patterns, we did not directly evaluate the review burden or cognitive load associated with scanning suggestions. Future work should incorporate controlled evaluations of latency, on-device or hybrid inference strategies, and a more explicit characterization of review cost, as these factors are critical for understanding how personalized AI scales to real-world AAC use.
\section{Positionality Statement and user profile}

The lead author of this work has been a full-time AAC user for over a decade, since a neuro-motor disease manifested when he was a teenager, affecting his ability to speak. He brings his lived experience navigating the complexities of communication asymmetry, authorship, and social presence in both personal and professional contexts. This dual perspective, as both researcher/developer and daily AAC user, deeply informs the framing, design, and analysis of this study. We disclose his experiences, as they played a significant role in shaping the research process, and contextualize the autoethnographic methodology.
\begin{figure}[h]
    \centering
    \includegraphics[width=1\linewidth]{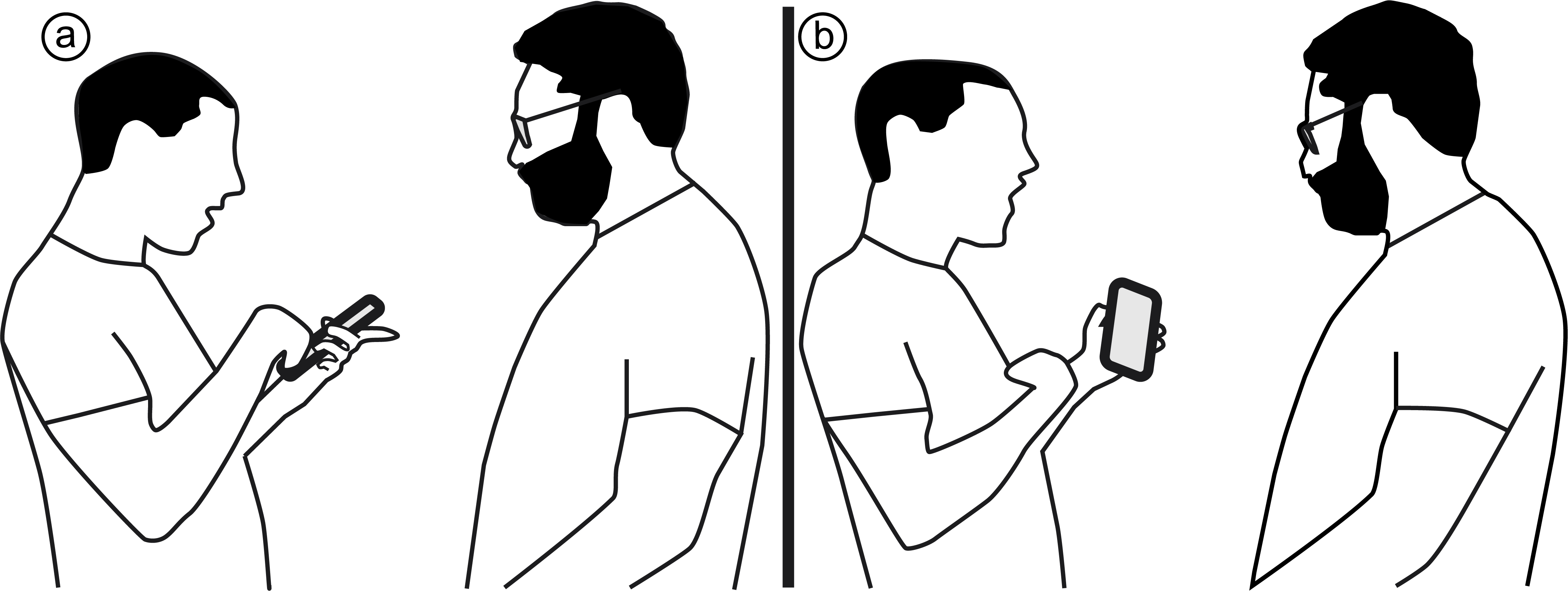}
    \caption{Shows the mode of conversational interaction: a) The lead author is typing a message. b) The lead author shows the message to his conversational partner.}
    \Description{Two-panel line drawing of an AAC conversation. (a) The lead author faces a partner while looking down and typing on a smartphone. (b) Separated by a vertical divider, the lead author holds up the phone so the partner can view the screen.}
    \label{fig:interaction}
\end{figure}
As shown in Fig. \ref{fig:interaction}, the author uses a “text-showing” style for most of his communication, typing in large fonts on his phone and showing the screen to interlocutors. This practice allows him to leverage the interlocutor’s imagination and his own non-verbal cues (e.g., facial expressions) to preserve tone and nuance, sidestepping the flat, unexpressive qualities of synthetic voices. 

These preferences also shaped the interface design. The phase 1 data collection app was built around minimal friction in note-taking, mirroring his everyday use of the iOS Notes app, before layering in personalized LLM suggestions. Thus, the design of the interface was not generic, but tuned to his own style and comfort with showing text. Beyond design, this perspective informed the analysis: his diary entries describe subtle experiential dynamics such as self-censorship when knowing conversations were logged, the difficulty of labeling interactions without disrupting flow, and the blurred boundary between system output and personal voice.

The role of user and developer brings a rather uncommon perspective where you are designing for yourself. We believe this adds to the experience of building an ultra-personalized system since it's not only the personalized LLM that enhanced the experience, but the exclusive design of the interface based on his communication style and needs.

We emphasize that the primary contribution of this work is not the technical system, but a multi-month empirical account in how an ultra-personalized, AI-powered AAC system affects everyday communication when trained on a single user’s AAC history and evaluated by that same user, focusing on its impact on agency, identity, and privacy through this lived experience. The author’s positionality as both user and researcher reveals how ultra-personalized AI can enhance fluency while also introducing new risks around privacy, authorship, and identity, where this first-person lens is essential for surfacing these tensions before naively deploying such technology in the wild.

\section{Related Work}

We build on work examining the challenges of expressivity and authorship in AAC, the trade-offs in AI writing support, and the implications of personalization, privacy, and agency in AI-mediated communication.

\subsection{Expressivity and Authorship in AAC}

AAC users typically compose messages at a rate (12–18 WPM), compared to typical speech (125–185 WPM)~\cite{waller_telling_2019}, creating what has been termed “communication asymmetry”~\cite{seale2020interaction}. This asymmetry can disrupt conversational rhythm, reduce opportunities for spontaneous expression, and create perceptions of disengagement~\cite{higginbotham2016time}. Yet timing alone does not determine communicative success. Research in AAC has shifted toward understanding \textit{conversational agency}, the capacity to express intent and interact in dynamic contexts~\cite{valencia_conversational_2020, barnlund1970transactional}. 

To address this asymmetry, researchers have explored tools such as word prediction~\cite{trnka_user_2009} and the use of visuals~\cite{fontana_de_vargas_aac_2022, fontana_de_vargas_automated_2021}. More recent systems move beyond these strategies by inferring suggestions directly from context, drawing on cues like location, conversation transcripts, or surrounding images~\cite{kane_lets_2017, kane_what_2012, valencia2024compa}. As AAC systems are increasingly integrated with generative AI to support faster communication, generic models often lack the nuance to reflect a user’s personal tone or identity~\cite{valencia_less_2023, shen_kwickchat_2022, weinberg2025why}. Concerns about authorship and control have also been raised: in a community consultation, AAC users and practitioners worried that AI systems could homogenize voices, misattribute authorship, and breach privacy through context-aware sensing, underscoring the need for safeguards and codes of practice~\cite{griffiths2024use}.

While supporting faster communication rates can be beneficial, personal expression remains a critical factor for AAC users. \citet{kane_at_2017} have shown the complex dynamics that AAC users go through to be able to express themselves and retain their identity. \citet{weinberg2025why} showed that in time-sensitive contexts, AAC users accept reduced agency in favor of delivering the comment faster.
Furthermore, by studying the backchanneling practices of AAC users, \citet{weinberg2025one} demonstrated that these unique cues form part of a broader AAC micro-culture, supporting self-expression and turn-taking. Together, these studies suggest that expressivity in AAC involves trade-offs negotiated in real time, shaped by social norms, contextual pressures, and individual communication goals.

Our work builds on these insights by investigating ultra-personalized AI systems trained on an AAC user’s own prior speech data instead of generic data. Unlike prior systems that offer speed at the cost of personal nuance, we examine whether an LLM fine-tuned on personal communication patterns can increase fluency without eroding the user’s identity. 

\subsection{AI Writing Support: Co-Authorship and Authenticity}

Advances in large language models (LLMs) have demonstrated the ability to generate human-like text~\cite{brown2020language}, adapt to individual writing styles~\cite{toshevska2025llm}, and assist in text composition~\cite{yuan2022wordcraft}.
Early work focused on predictive text entry~\cite{bi2014complete, vertanen2015velocitap}. More recent systems move beyond token-level prediction toward sentence- and paragraph-level assistance, but with distinct aims: for example, tools for ideation generate prompts for the user to inspire new directions for science writing~\cite{gero2022sparks}, while post-hoc revision interfaces support correction and editing~\cite{cui2020text}. Work on human–AI co-writing examines how model contributions interact with writers and writing outcomes:  \textit{CoAuthor}~\cite{lee2022coauthor} introduces a dataset of assisted writing sessions to study collaboration dynamics in creative and argumentative writing, while other domain-specific systems \textit{Dramatron}~\cite{mirowski2023screenplays}, explore hierarchical prompting for long-form co-creativity screenplay writing with industry evaluations. Furthermore, controlled experiments show that opinionated assistants can shape both what users write and what they subsequently believe~\cite{jakesch2023co, fu2023comparing, agarwal2025ai}.

 Researchers have explored how AI tools can scaffold writing by visual prompts~\cite{chung2022talebrush} and by text summarization~\cite{dang2022beyond}. Others investigate AI’s role in collaboration and social dynamics, showing how suggestions affect group processes or perceptions of ownership~\cite{gero2023social, stark2023can}. A different line of work focuses on stylistic and genre-specific support, such as generating metaphors~\cite{kim2023metaphorian}. More recent systems, like \textit{WordFlow}~\cite{wang2024wordflow}, highlight how collaborative prompt design can enhance the writing process. Finally, systematic reviews synthesize these developments, mapping how AI writing tools are used across diverse tasks and user groups~\cite{fang2023systematic, lee2024design}.

  While these systems promise increased productivity and fluency, they may also reshape a user's tone, intent, or perceived authorship~\cite{kadoma2024role}. Recent empirical work shows that people often do not perceive themselves as owners/authors of AI-generated text, yet still self-declare authorship, what Draxler et al. term the “AI Ghostwriter Effect ”~\cite{draxler2024ai}, and that personalization alone does not resolve this effect; perceived ownership increases mainly with greater human influence over the output. In contrast, \citet{mishra2025chat} found that LLMs in collaborative writing can produce a sense of chat-ghosting, where system contributions overshadow human intent and leave participants feeling displaced from authorship. As LLMs become more powerful, the use of AI writing assistants raises longstanding questions about co-authorship, intent, and authenticity. At the same time, such systems often operate as opaque, one-directional suggesters offering completions or rewrites without revealing how they interpret the user’s intent. The complexity of articulating a prompt can elicit feelings of limited agency when working with LLMs \cite{ippolito2022creative,   zamfirescu2023johnny} subtly steering expression, masking alternatives, or requiring users to “write against” the system.

This problem is magnified for users with limited physical input or slower composition rates: accepting a suggestion may be faster, but rejecting or editing it often requires disproportionate effort~\cite{valencia_less_2023}. \citet{valencia_conversational_2020} found that AAC users were wary of being perceived as lazy or inauthentic when relying on AI-generated responses. Similarly, in \textit{Why So Serious?}, some users described discomfort when AI responses blurred the line between their own voice and system-generated output~\cite{weinberg2025why}. Complementing design-centric work, ethics analyses warn that LLMs in VOCAs can blur authorship and remove co-construction opportunities, arguing against technocentric adoption and for explicit attention to humanness and user-centred design~\cite{griffiths2025ai}. These findings call for tools that support user authorship without compromising responsiveness.

Our work extends this conversation by exploring what happens when the AI model is trained not just on general data, but on a user's own communication history. We show how deep personalization can enhance fluency and relevance, while also introducing new forms of entanglement between user identity and system output.

\subsection{Privacy and Personalization in AI-Mediated Communication}

Personalization in AI-mediated communication promises fluency and relevance, but also raises concerns about authorship, privacy, and control. \citet{chen2019gmail} demonstrates large-scale AI-assisted writing and discusses deployment and fairness considerations in a personalized feature, while \citet{hermann2022artificial} emphasizes the need for AI literacy and multi-stakeholder oversight to mitigate ethical risks.
Frameworks like Contextual Integrity~\cite{nissenbaum2004privacy} define privacy as the preservation of appropriate information flows between stakeholders, where violations occur when information shared in one setting is reused in another without regard for the original norms and expectations. In AAC, this manifests when ultra-personalized systems resurface private expressions in public or professional settings, even without external data sharing.

Recent systems frame personalization as a dynamic process rather than a fixed model state. One system supports user control through adjustable stylistic boundaries, showing that interpretability and real-time negotiation can help preserve authorship~\cite{yeh2024ghostwriter}. Others highlight how design decisions shape the user’s voice, through scaffolded generation pipelines~\cite{li2023teach}, explicit cues to preserve tone~\cite{peng2024reviewllm}, or lightweight embeddings that inject style while avoiding full retraining~\cite{liu2024llms+}, which outperform existing methods on the LaMP benchmark~\cite{salemi2024lamp} without requiring parameter tuning. These approaches point to a shift: personalization is no longer just a tool for fluency; it is a question about identity, authorship, and agency. In AAC, where generated text becomes the user’s literal voice, this negotiation plays out word by word.

Personalization does not just affect output quality; it shapes how users relate to AI over time. As systems become more responsive, users begin to ascribe trust, intent, and even social presence to them, with these perceptions shifting based on context and relational closeness~\cite{araujo2024speaking, lim2022no}. These dynamics raise ethical concerns around transparency, control, and the limits of consent, especially when personalization becomes invisible or automatic~\cite{king2024rethinking}. In AAC contexts, where AI-generated text functions as a user’s voice, these stakes are amplified: trust and privacy are fundamental for expressing identity safely and authentically.

While our focus is not on securing data pipelines, privacy-preserving personalization has been widely explored in parallel fields. Techniques such as federated learning, on-device adaptation, differential privacy, and secure aggregation offer important infrastructural safeguards for real-world deployment~\cite{mcmahan2017communication, bonawitz2017practical, hard2018federated, abadi2016deep}. These methods address data protection at the system level, but they do not resolve the lived, moment-to-moment tensions around visibility, consent, and contextual misuse that emerge when personalized AI becomes part of everyday communication. Our work complements these efforts by examining how such tensions unfold experientially, particularly when the system functions as the user’s voice.

 Work in value-sensitive design~\cite{friedman2013value} and user-controlled privacy~\cite{erlingsson2014rappor} has argued for local differential privacy, but rarely in the context of everyday conversation. In AAC, where authored text is not only logged but literally becomes the user’s spoken voice, the stakes are higher.

Our autoethnographic study exposes these stakes in lived experience. We examine privacy–personalization tensions not in abstract or technical terms, but in the moment-to-moment dynamics of AAC speech. This perspective reveals that ultra-personalization must account for context, relationship, and the shifting boundary between the user’s and the AI’s voice.

\section{Phase 1: Data Collection}
We started out by collecting seven months of in-person conversation data of the lead author. In this phase, we describe the autoethnographic methodology used, the design of the data collection tool, and conclude with reflections on the first-person observations. 
\begin{figure}[b]
    \centering
    \includegraphics[width=0.9\linewidth]{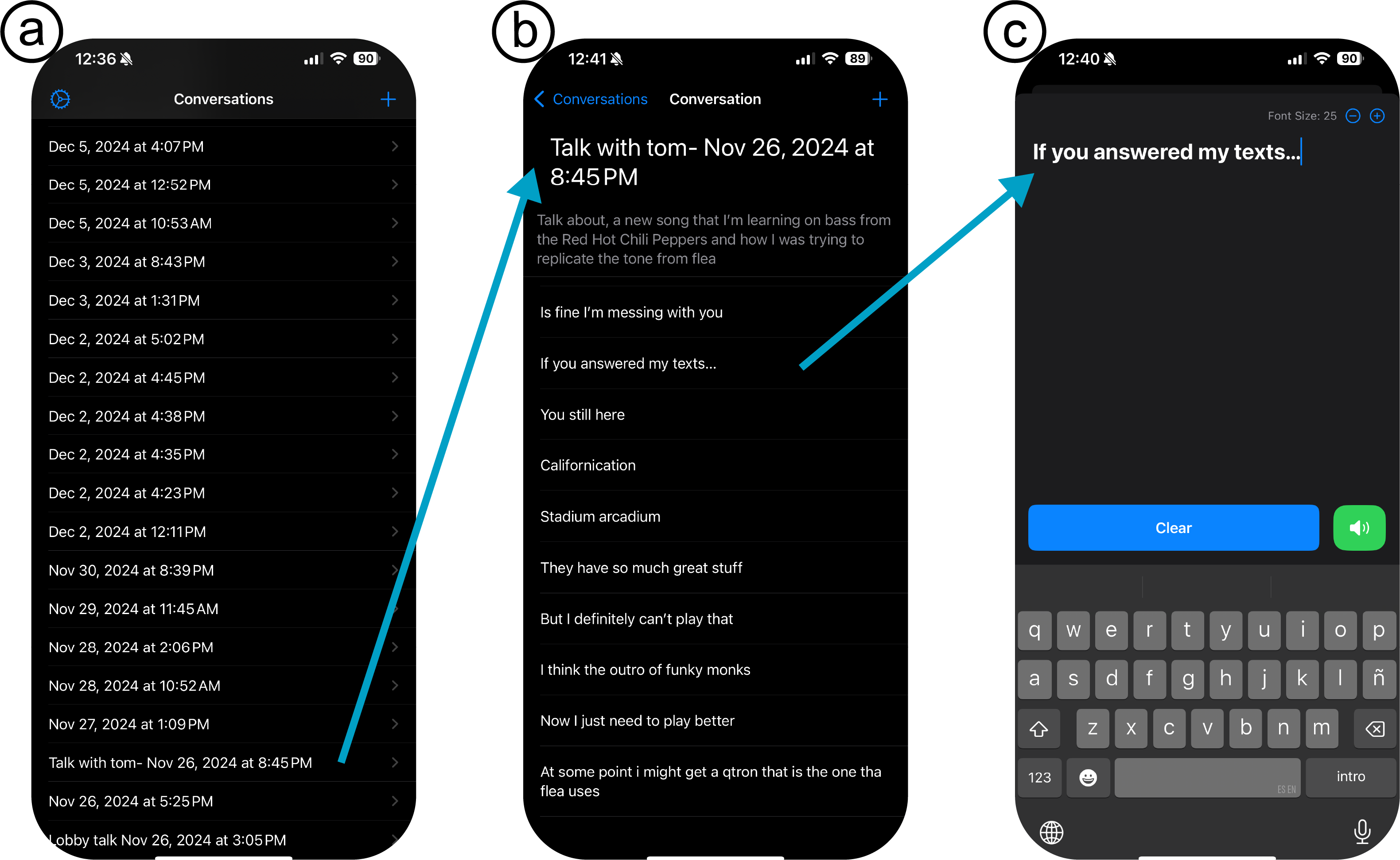}
    \caption{Shows the workflow of using the app. a) landing screen when you open the app, which shows the conversation threads where you can create a new one or select a previous one. b) an active thread, shows the title and context description (editable) and past messages, and you can create new messages to the active thread. c) composer modal, where the users would write and display the text, with controls for font size on top. The clear button would append the message to the active thread and clear the text, speak button reads the text out loud.}
    \Description{Four mobile screenshots show the app workflow. (a) ‘Conversations’ list with dated threads and a plus button to create a new one. (b) A selected thread titled ‘Talk with Tom, Nov 26, 2024, 8:45 PM’ with an editable context line and past messages. (c) Composer view typing ‘If you answered my texts…’, font-size controls, a Clear button, and a Speak button. }
    \label{fig:interface_phase_1}
\end{figure}
\subsection{Methodology}

Over a period of seven months (November 2024 – May 2025), the lead author used a custom-built iOS app as his primary in-person communication tool. Every face-to-face communication was typed using the app, allowing it to act as both an AAC interface and a data collection platform.
During this time, the lead author maintained reflective notes in the form of open-ended entries to document his experiences during the data collection phase. The app continuously logged only the lead author’s side of face-to-face communication. 

We experimented with several ways of delineating conversations, grouping by physical context and by conversation topic, but ultimately, the only trackable approach was segmenting by day, which provided a consistent and manageable unit of analysis, at the cost of topical granularity. These methodological choices, and the reflections they prompted, shaped the dataset into a form that was authentic to lived AAC use and tractable for subsequent analysis.

Conversations data was accompanied by optional annotations describing the context and communicative intent (entered by the lead author). These annotations included a short textual description of the topic and setting for conversations.

\subsection{Data Collection Interface Design}

The data collection interface was designed to be similar to the iOS Notes app (the author's previous mode of communication) to be minimally invasive in his communication workflow, while providing a systematic way of collecting the data. The app was developed in Swift and deployed on an iPhone 16 Pro Max.

As shown in Fig. \ref{fig:interface_phase_1}a, the first screen presents a list of conversation threads. Tapping the \texttt{+} button starts a new conversation and opens the text editor for the lead author to type. Selecting an existing conversation opens its message history for that conversation Fig. \ref{fig:interface_phase_1}b which has a title and a context description that the lead author can edit to add relevant information to that particular conversation, and also the lead author can review past messages or tap \texttt{+} to add a new message to that conversation, which opens the text editor. 

\subsection{Raw Data}
The Phase 1 data collection app structured data into conversation threads, where each thread acted as a parent node containing multiple messages. Every thread included a title and a short description entered by the lead author to capture contextual information (e.g., environment, or interlocutor). Each message stored within a thread contained the text content, its parent thread ID, and timestamp.

 \begin{figure}
    \centering
    \includegraphics[width=1\linewidth]{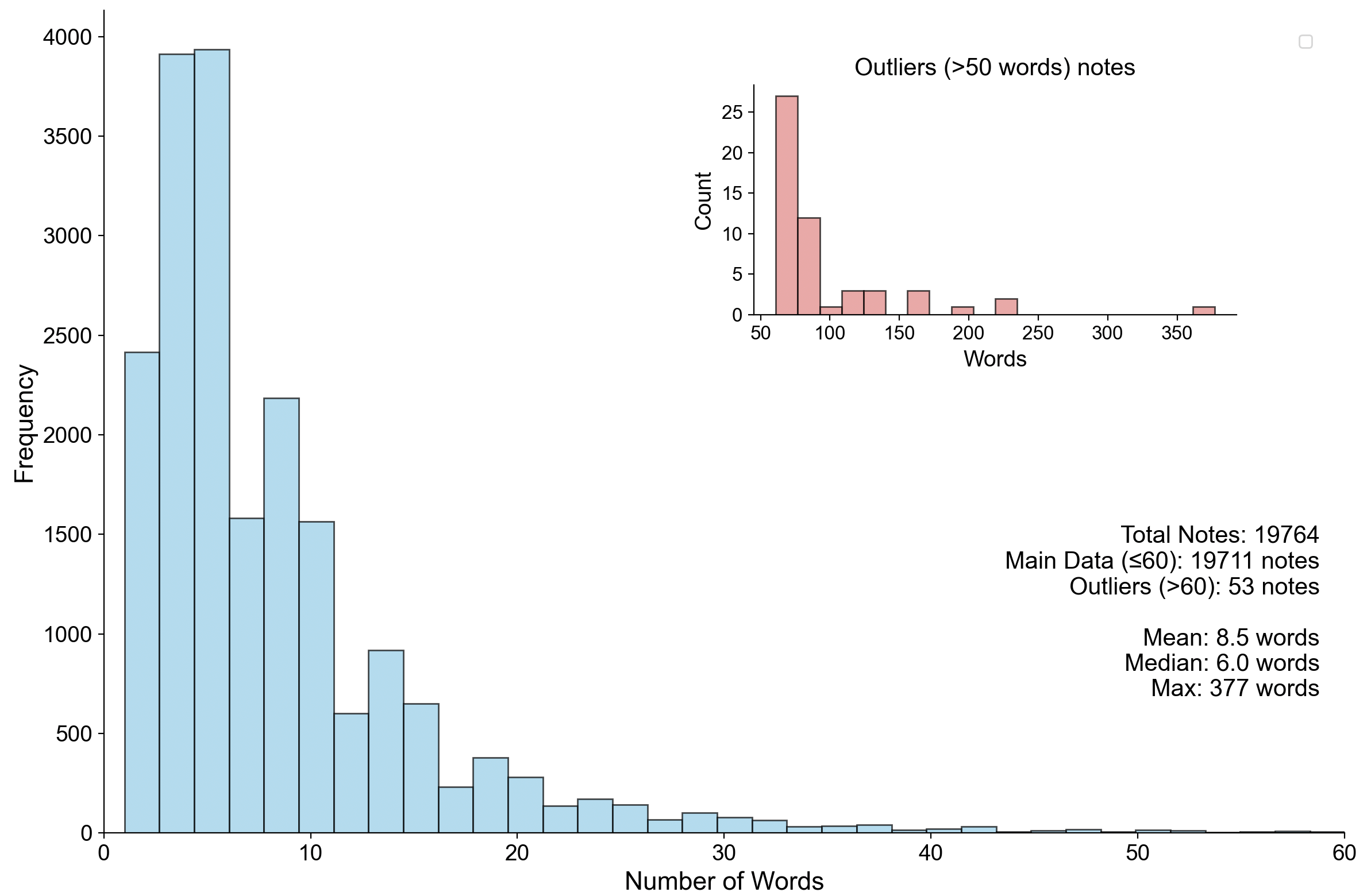}
    \caption{Distribution of word count in messages.}
    \Description{Histogram of message word counts (x-axis: Number of Words 0–60; y-axis: Frequency). The distribution is strongly right-skewed, most messages are short, peaking around 3–10 words, with frequencies decreasing steadily past ~12 words and a long tail toward 50–60. The inset shows outliers >50 words (53 notes), with a few reaching up to 377 words; total notes = 19,764, mean = 8.5 words, median = 6.}
    \label{fig:distribuition_phase_1}
\end{figure}

Over the course of 7 months, we collected 247 conversation threads spanning 19,764 messages. These threads captured a wide range of communicative contexts, ranging from professional discussions to casual conversations in bars and even brief interactions such as ordering drinks on a plane. Fig. \ref{fig:distribuition_phase_1} shows that most messages are relatively short (mean 8.5 words) responses in conversations with some outliers where the lead author explained more of a monologue at once, for example, a lab meeting update with 377 words.
\\ 
\\
An example data unit is shown below:
  \noindent

  \begin{lstlisting}[ basicstyle=\ttfamily\small]
  {   
   "title": "Maker space",
   "description": "Went to makerspace to ask 
   for bubble wrap to ship a graphics card",
   "id": "D7DAA372-BAE5-49D1-924A-A79F9C856917",
   "notes" : [
     { "content": "Do we have bubble wrap ?",
       "conversationId": 
       "D7DAA372-BAE5-49D1-924A-A79F9C856917",
       "humanReadableTimestamp": "2024-11-25 16:08:58",
       "id":
       "E1D19C76-AA05-48E0-897E-E8A7DE0754C4",
       "timestamp": 754261738.067545},
     { "content": "Can I borrow the cart for a bit?",
       "conversationId": 
       "D7DAA372-BAE5-49D1-924A-A79F9C856917",
       "humanReadableTimestamp": "2024-11-25 16:12:27",
       "id" : "9E0BA377-83F9-47DA-8120-C30C8A001497",
       "timestamp": 754261947.620366} 
   ],
  },
    \end{lstlisting}

 \subsection{Diary Reflections} 
This section presents both the personal reflections on how data collection shaped the author’s communication over the course of seven months, and as a methodological account of the practical challenges that shaped the dataset.
We summarized and categorized the diary entries into 3 themes: (1) \textit{Speaking Under Observation: How Logging Constrained My Agency}, (2) \textit{Constructing a Learnable Identity: Boundaries, Segmentation, and Representation}, and (3) \textit{Context Sensitivity Risks Privacy Breaches}. To reflect the personal nature of these observations, we report these from a first-person perspective.

\subsubsection{Speaking Under Observation: How Logging Constrained My Agency}

One of the earliest and most striking observations during the data collection process was the profound impact that constant logging had on my communication habits. In the first two weeks, I noticed a gradual but noticeable increase in self-censorship, where I consciously avoided certain words, phrases, or tonal choices that I would otherwise use freely in everyday conversations. For example, using “F**k” in a sentence, or gossiping about someone in the building with the doorman, or even making some dark humor jokes as I would normally do. This change was not a deliberate effort to alter my communication style; rather, it was an automatic, subconscious response to knowing that my words were being stored. 

What became immediately apparent was how logging changed my sense of agency over my own speech. Instead of expressing myself freely as I naturally would,
I became acutely aware of how I phrased things, much like a person who refrains from swearing in front of children or adjusts their speech when they know they are being recorded. 
This awareness created a form of self-monitoring not typically present in spoken interaction, constraining my agency to speak naturally. Routine exchanges, like talking to a girl at a bar, hanging out with my colleagues in the office, or late-night existential talks with a close friend, took on a different weight when preserved as text logs, raising questions about how knowing that speech is being logged shapes what users choose to express.

At times, I withheld expressions that felt authentic simply because I did not want them to be learned, repeated, or resurfaced later by the personalized model. There was one time when I reverted back to my old IOS notes app because I needed to complain about a neighbor with a friend, and since I was going to say "not very nice things" about this person with whom I had an ongoing dispute, I didn't want to have any records of what I said.
I found myself managing what would become part of the dataset, choosing which aspects of my voice I was comfortable having represented and which I preferred to withhold. 
Instead of typing solely for the immediate interaction, I found myself typing with an awareness of how each utterance might contribute to a lasting representation of my identity. This constrained the freedom and spontaneity that typically characterize everyday talk, making self-censorship and shifts in communication style more likely over time.

\subsubsection{Constructing a Learnable Identity: Boundaries, Segmentation, and Representation}
Spoken language is inherently transient; it exists momentarily before fading away, leaving no permanent record. While AAC text, especially when logged, acquires a kind of permanence. During Phase 1, this permanence became particularly salient to me: I sometimes noted avoiding certain words or jokes with close friends because I found myself thinking, "I don’t want the LLM to learn this from me.” I realized that expressing them would not just communicate something in the moment, it would also shape the version of my identity that the model would later learn and reproduce. What would otherwise be an ephemeral, low-stakes comment felt instead like contributing to a persistent representation of myself that I could not fully control. This introduced a layer of identity management into otherwise ordinary moments.

Another challenge I faced during data collection was determining when a conversation truly ends and when a new one begins. This issue became particularly apparent when trying to split conversations into discrete units for analysis. For instance, if I am having a conversation with one of my labmates in the lab and then transition to a discussion with another, the question arises: Is this a continuation of the same conversation, or does it represent the start of a new one? Does the topic change affect this distinction, or is it more about the social context in which the interaction occurs?
This ambiguity mattered for analysis and also for how I understood the identity that the dataset would ultimately encode. Each segmentation decision effectively determined which interactions “belonged together,” shaping what the model might later infer about my tone, relationships, and style across contexts.

This segmentation dilemma was further complicated when I attempted to group conversations based on location. For example, when I went to a bar to play music, I found myself speaking with several different people in the same setting, yet the topics of discussion varied widely. One moment, I might be talking about the chords we were playing, and the next, I could be asking about a friend’s father’s health. These exchanges might be followed by a philosophical debate with yet another friend. This approach to grouping conversations by location created an overly large thread, containing diverse topics ranging from mundane to deeply personal. In this case, the contextual setting (the bar) did not provide sufficient cohesion between the topics, resulting in a jumble of unrelated discussions. 
Treating all of these as a single “conversation” highlighted how segmentation choices could collapse very different facets of my communication into one undifferentiated block. This made it clear that fine-grained boundaries matter for understanding how identity and tone shift across contexts, even though such distinctions were difficult to capture in real time. The ambiguity of conversational boundaries without clear segments, context-dependent remarks risked being logged in ways that no longer matched their intended social setting.

Real-time segmentation added a layer of conversation management quickly became untractable in my everyday life, taking my attention away from the conversation itself and thus affecting my interpersonal relationships. Therefore, I adopted a simpler and more reliable approach: conversations were delineated by day, more akin to diary entries than discrete sessions. This ensured that all face-to-face communication within a given day was preserved without the burden of making fine-grained segmentation decisions in real time. While this method sacrificed topical granularity, it provided a consistent, low-overhead way of structuring the data while still allowing for later reflective analysis.
This approach prioritized feasibility during data collection, understanding that more nuanced conversational boundaries could be analyzed retrospectively.

In practice, data collection required me to manage both what I wanted to say in the moment and how I wanted my identity to be encoded for the personalized model, complicating the freedom and immediacy that normally characterize everyday talk.

\subsubsection{Context Sensitivity Risks Privacy Breaches}

The need for vast amounts of data to effectively train ultra-personalized LLMs creates another layer of concern. Speech is highly context-dependent, meaning that what is appropriate or acceptable in one setting may be entirely inappropriate in another. A casual inside joke with a close friend, for example, “We basically get paid to f**k around” humorously referring to our job as researchers, which might be humorous in the context of friends/co-workers, but would be totally inappropriate or offensive in a different setting, e.g., a talk with your advisor about research funding.

Drawing from the \textit{contextual integrity}~\cite{nissenbaum2004privacy} concept, the flow of information in a different context is only appropriate if the usefulness of that information outweighs the integrity violation. In practice, humorous or playful remarks often have little informational value; reusing them outside their original context would feel inappropriate and potentially breach privacy.

This meant that even before any model was trained, the act of logging and aggregating conversations already produced a sense of privacy risk. Without clean boundaries, remarks meant for one relationship or moment became folded into a larger dataset whose internal logic no longer matched the social logic that gave those remarks their meaning. A joke intended only for a close friend, a complaint meant for a single confidant, or a casual remark shaped by the mood of the moment all became side-by-side entries in a single running log. As soon as they were preserved together, their original contextual “containers’’ dissolved, making it difficult to tell which utterances still belonged to which parts of my social world.

The loss of contextual separation was itself a form of exposure: not because others could see the data, but because the data no longer clearly reflected the relationship boundaries and situational cues that normally govern what feels safe to say. Once aggregated, these utterances took on a second life as decontextualized text, detached from the interpersonal dynamics that had made them appropriate, and this shift subtly eroded my sense of control over how my voice and intentions were represented.

\section{Phase 2: Curation and Model Training}

In this section, we describe how we prepared the training data and the steps we followed to train our ultra-personalized model. This process included four major parts:  (1)~filtering data, (2)~ structuring the data for training, (3)~ training a Personalized Model (trials with different models, and testing different training hyperparameters), and (4) augmenting the interface with the trained model.

\subsection{Filtering Data}

We curated a subset of the collected data to train a personalized language model. We first started by filtering messages that the lead author did not want to publicly disclose. We applied a standard toxicity filtering pipeline~\cite{logacheva2022paradetox} to exclude any messages containing potentially offensive or inappropriate content. While this filtering step inevitably reduced the “truthfulness” or fidelity of the raw dataset, it removed utterances that may have reflected authentic but messy moments of interaction.
This was especially important given that the model was trained using OpenAI’s fine-tuning API, which required uploading data to external servers.

The decision to filter the data was directly shaped by insights from Phase~1: the same kinds of utterances that felt risky or overly revealing when logged were the ones we chose to filter out when preparing the training data. In this sense, filtering became a second opportunity to reinstate the conversational boundaries that were difficult to maintain during real-time logging.
Beyond privacy, this filtering also reflected a deliberate design choice about authorship. The lead author did not want the AI to be responsible for producing utterances such as profanity, curse words, or dark humor. Although these play a role in everyday communication, he did not trust the model to generate them in a way that was contextually appropriate. By excluding them from training, he retained his agency over when and how these more sensitive or risky expressions occur. In practice, this meant that the model was shaped to produce only the kinds of utterances the lead author was comfortable delegating to it, while reserving sensitive forms of speech for his own discretion. In this sense, filtering was not just a technical preprocessing step, but a way of aligning the model with the lead author's communicative boundaries, and maintaining his agency.

Out of 19,764 messages, 269 were filtered out automatically by the toxicity pipeline. Examples of such filtered utterances include: “It’s not just bull***t I wrote.” These removals illustrate the kinds of expressions that, while authentic to everyday communication, were judged unsuitable for inclusion in the training data, both because of privacy concerns and because the author did not want the model to reproduce them automatically.

\subsection{Structuring Data for Training}

To fine-tune LLMs, training data must be formatted to follow the model’s specific input–output conventions. Beyond formatting, the data also needs to be structured so that the model can “learn” when and what tokens to generate given an input (i.e., modeling the probabilistic distribution of word sequences encoded in the training data). For example, our final model was a fine-tuned version of GPT-4.1-mini. Preparing a training corpus for this model required reformatting each conversational entry into paired training examples consisting of a user message and an assistant message, which in normal instruction LLM training would be pairs of questions and answers data. In our case, the user message corresponds to text written by the lead author, while the assistant message represents the expected continuation that the model should attempt to predict. We applied this reformatting consistently across all models we evaluated, adjusting to each model’s specific data requirements.

We adopted a simple incremental chunking approach: splitting each message by word, predicting the remaining portion, and sequentially adding more words until the $N-1$th token. For instance, the sentence “hello, how are you?” would be transformed into the following dataset examples:

\begin{lstlisting}[ basicstyle=\ttfamily\small]
{messages: [{"role": "system", "content": "<SP>"},
{"role": "user", "content": "hello,"},
{"role": "assistant", "content": "how are you?"}]},

{messages: [{"role": "system", "content": "<SP>"},
{"role": "user", "content": "hello, how"},
{"role": "assistant", "content": " are you?"}]},

{messages: [{"role": "system", "content": "<SP>"},
{"role": "user", "content": "hello, how are"},
{"role": "assistant", "content": " you?"}]},
\end{lstlisting}

The training \texttt{<SP>} = system prompt used in all examples can be found in the appendix.

This incremental splitting method allowed us to augment the dataset size while aligning with the model’s training dynamics. In practice, it proved effective at deliberately overfitting the model to our task, ensuring that the fine-tuned system captured the author’s idiosyncratic communication patterns.

\subsection{Training a Personalized Model} \label{subs:trainingpersonalized}

The personalized model was fine-tuned on the author’s conversational corpus collected during Phase 1. We evaluated several candidate LLMs, ranging from smaller models that can run on-device, such as Gemma 3 and DeepSeek-R1-Distill-Qwen-7B, to larger models cloud-based, such as Llama 3.5, Cohere's Command-R-08-2024, and GPT-4.1-mini. In conventional machine learning practice, the primary metric for model selection is accuracy. However, for ultra-personalized systems, accuracy alone proved insufficient. What mattered most was whether the model’s output felt representative of the author’s style and cultural-linguistic background.

This criterion of alignment with self-representation is inherently subjective, but it reflects the central goal of personalization: to preserve the user’s sense of authorship and identity. For example, even when smaller models performed reasonably well on standard benchmarks, their completions often felt generic or failed to capture the author’s bilingual (English/Spanish) communication style. Multilingual capability thus became a second decisive factor, narrowing the viable candidates to Gemma, Llama, Cohere, and GPT.
Through direct testing with comparable completion prompts like "hello, I'm " or "my research is " across several regenerations, the lead author voted on which one was similar to what he would write. Gemma was discarded because, after fine-tuning, the coherence of the output sentences significantly degraded, and Llama 3.5 gave too generic answers; the final decision boiled down to Cohere vs GPT-4.1-mini, where GPT-4.1-mini emerged as the model that best reflected the author’s communicative voice across languages and contexts. It was therefore selected and implemented in the app.

% \TODO{cohere we had 38\% accuracy, and 41\% accuracy, with GPT we got 80+\%}

% This decision underscores a broader methodological point: while accuracy remains important, ultra-personalized AAC systems the user’s subjective experience of recognition and representation greater weight. Ultimately, it is this alignment with the user’s lived voice, not benchmark scores, that determines whether a model is usable and empowering.

\begin{figure}[h]
    \centering
    \includegraphics[width=1\linewidth]{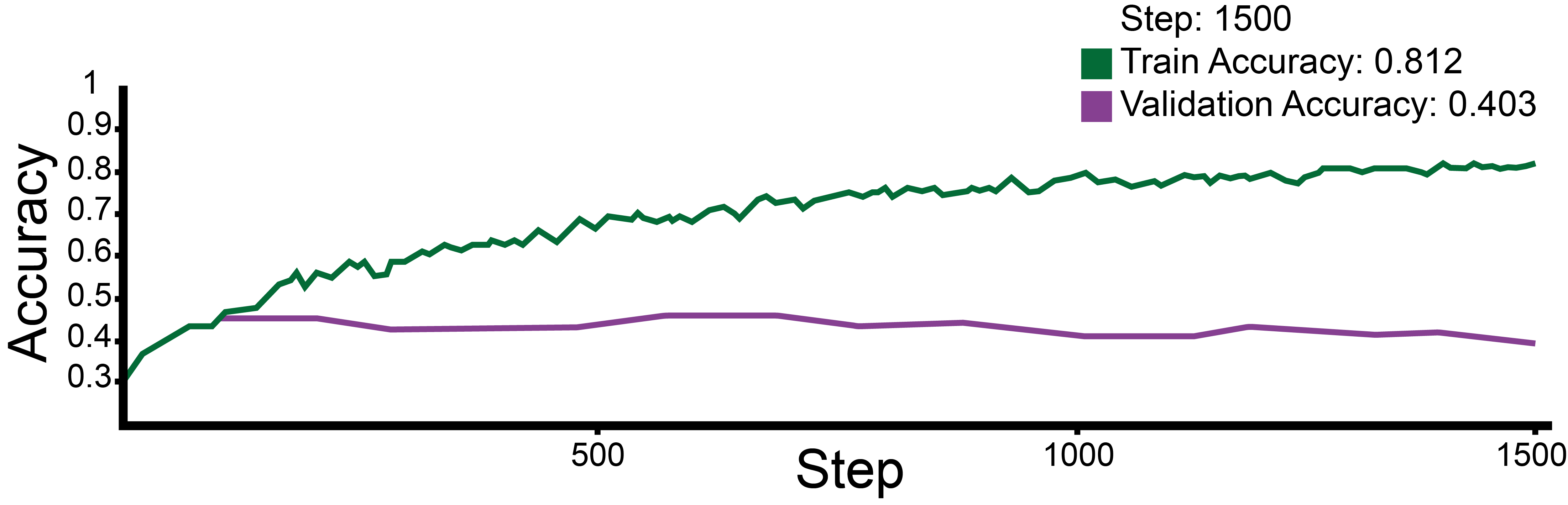}
    \caption{OpenAI Platform training/validation accuracy curve (smoothing: 0.85).}
    \Description{Line chart of training accuracy over steps (x-axis 0–1500; y-axis 0–1). Training accuracy rises from ~0.30 at the start to ~0.81 by step ~1500, increasing steadily with small fluctuations; Validation accuracy stays almost constant at ~0.40. Curve shown with smoothing 0.85.}
    \label{fig:accuracy}
\end{figure}

GPT-4.1-mini was fine-tuned using supervised fine-tuning (SFT) within the OpenAI platform service, which enables semi-automatic training with custom examples tailored to specific use cases. The resulting model reliably produced text in the author’s desired style and tone. The dataset order was randomized and split using standard 80/20\% for train/validation sets, training was performed on 22,415,365 tokens over one epoch, with a batch size of 64 and a learning rate multiplier of 2, OpenAI platform does not provide any additional information about the hyperparameter tuning that is done internally. This configuration resulted in a training loss of 1.087 and a validation loss of 4.889, corresponding to a training accuracy of 81.2\% and validation accuracy 40.3\% (Fig. \ref{fig:accuracy}). The total cost of training using OpenAI services was approximately \$120.

\subsection{System prompt}
In order to generate appropriate completions from the fine-tuned model, the system prompt that goes along with the request was crafted to reflect the lead author's intent. To add flexibility, the lead author added a way to customize the prompt from the settings screen of the app. Here we present the final version of the prompt that was a modification of the training prompt after using it.
\\
\begin{lstlisting}[ basicstyle=\ttfamily\small]
You are a highly personalized autocomplete
assistant designed to help Tobi,
who is a user with a speech disability, communicate
faster. You have been trained exclusively on this
user's past conversations and writing. Your role 
is to help them complete their thoughts naturally,
based on how they usually speak. Use vocabulary,
tone, and phrasing that match the user's unique
communication style. Do not introduce new ideas or
unfamiliar phrasing. Your goal is to continue the
user's message exactly as they would.  Wait for
the user to start a message, and then complete 
it in a way that sounds like them. If a message 
seems like a joke or sarcastic comment, complete
it with their typical humor style. If a message
is practical or neutral, complete it helpfully
and in their usual tone. Never include greetings,
explanations, or filler unless the user began
with one.
\end{lstlisting}

An important modification made to this prompt after the lead author started using the assistant was adding the  \textit{"to help Tobi who is a user with a speech disability"}, giving the specific name of the user made the LLM be able to generate more accurate introductions for the user.

\subsection{Interface AI augmentation}

After the LLM training, we extended the app with the ultra-personalized LLM that proposes short suggestions during message composition. The goal was to preserve the display-first workflow and agency established in phase 1 while offering lightweight assistance that feels like the author’s voice. Additionally, we added a toggle button at the top to enable/disable the LLM suggestions.

\begin{figure}[h]
    \centering
    \includegraphics[width=0.4\linewidth]{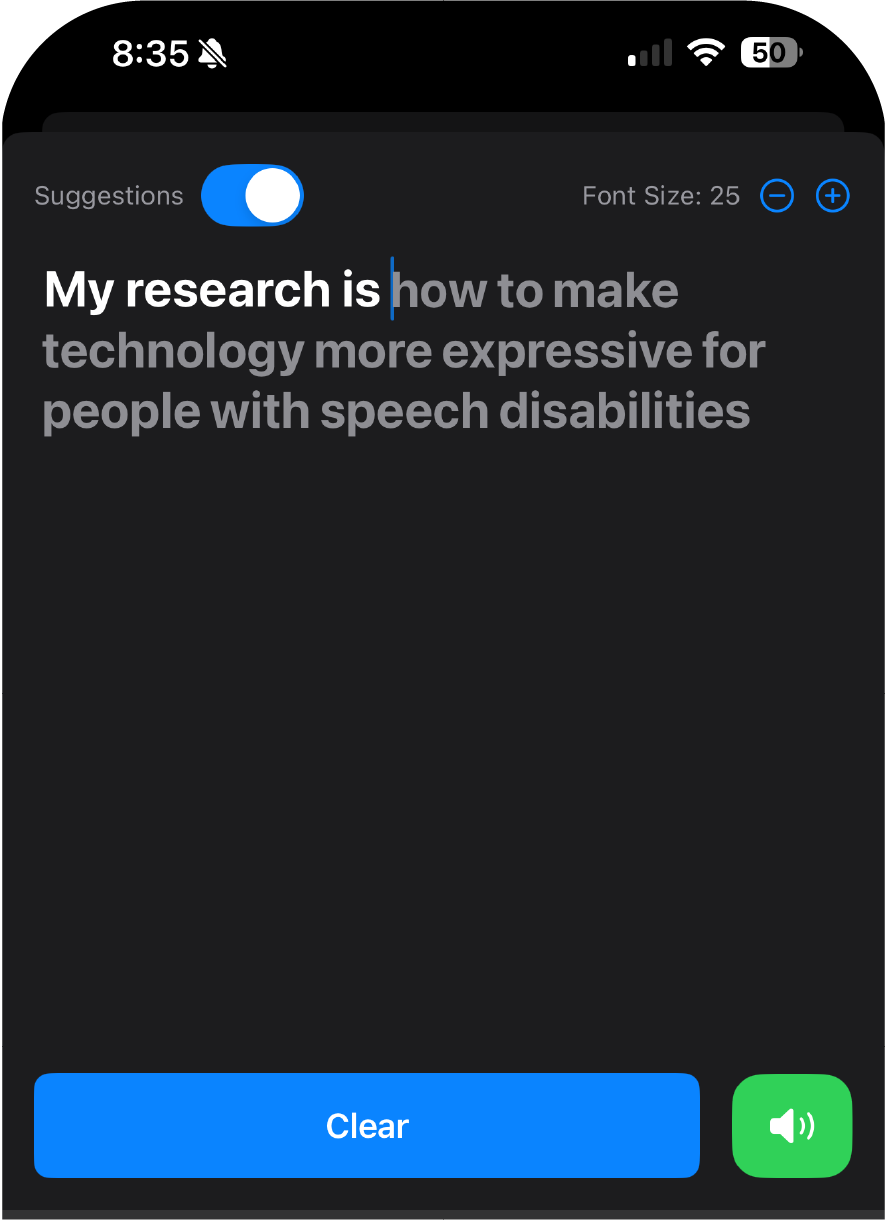}
    \caption{Example of text completion from the LLM.}
    \Description{"Mobile app screenshot showing LLM completion. With ‘Suggestions’ toggled on and font-size controls visible, the editor displays “My research is” (typed text) followed by a lighter grey suggestion: “how to make technology more expressive for people with speech disabilities.” A wide Clear button and a green Speak button appear at the bottom.}
    \label{fig:example_completion}
\end{figure}

Users request suggestions through semi-explicit interaction: pressing the space bar at the end of a word in the text editor triggers a cancellable request for a short next-phrase completion. Any subsequent keystroke immediately cancels in-flight requests and hides the suggestion. Consequently, nearly every typed word triggered a call to the LLM, making the suggestions "always" available without an explicit request, similar to how the word prediction of IOS works; practically, this means that most suggestions remained unused by design. For this reason, we did not explicitly track “rejected” suggestions, as that count would closely approximate the total number of typed words and would not provide a meaningful signal in our analysis.
Importantly, the model was not provided with any conversational history beyond the text typed in the current turn. Each suggestion was generated solely from the partial message the user was composing, without access to prior utterances. We adopted this single-turn approach because Phase 1 revealed that we could not reliably determine where conversations began or ended across real-world interactions. Given this ambiguity, we took a pragmatic design path: treating each turn as an independent unit and avoiding multi-utterance context that would have required consistent, accurate segmentation of conversations. Fig.~\ref{fig:example_completion} shows how suggestions are shown as \emph{inline ghost text} directly after the caret. They do not modify the committed text until the user accepts with a tap. Continuing to type simply overwrites the ghost text.

To preserve user agency, we implemented a partial suggestion acceptance mechanism similar to \cite{zindulka2025exploring, arnold2016suggesting} that allows more agency over how to accept the suggestions, by letting the LLM run free with generation length and giving the user the agency of choosing until which word of the suggestion to accept. This interaction allowed the user to benefit from full paragraphs (up to 55-word suggestions) when the LLM generated accurate suggestions, and also benefit from just a few-word completion that complemented the user's writing. We believe that this mode of interaction, combined with IOS keyboard one-word suggestions, equipped the user with maximum agency over their words while benefiting from the generative AI's full capabilities.

\section{Phase 3: Deployment and Evaluation}
In this phase, we deploy the application for three months. We describe the methodology used in this phase, and report both usage analytics and qualitative first-person reflections of the lead author.

\subsection{Methodology}

After training, we deployed the app on the lead author's iPhone 16 Pro Max for daily use over three months (June – August 2025). The app integrated the personalized model to provide AI-generated in-line suggestions in real time as the lead author composes messages.

During this period, the lead author completed structured diary entries regularly (at least once every three days) using a survey that combined Likert-scale ratings and open-ended reflections. The lead author first indicated whether AI-generated suggestions were used that day (frequently, occasionally, or not at all). They then rated the helpfulness of suggestions on a 5-point scale (1 = not helpful, 5 = very helpful) and described one concrete moment where a suggestion worked well or poorly. Additional items asked whether the LLM reflected or adapted to the author’s personality, whether more or less personalization would have been preferred, and whether any surprising or inappropriate phrases or tones appeared.

Agency, identity, privacy were explicitly addressed through targeted items: he rated his feeling of control of how the app represented his "voice" that day (1 = rarely, 5 = always), and whether any suggestions felt too personal or unexpected. Finally, he reflected on any discomfort about how his data might be used. Each diary entry concluded with an open-ended field for any additional reflections. 

The lead author and a second researcher conducted two rounds of affinity diagramming on the diary entries, first clustering raw observations into preliminary groups, then iteratively consolidating and abstracting these clusters into higher-level themes through discussion and consensus. 

In parallel, the app logged interaction data, tracking the following metrics:  duration of the interaction, the number of words typed by the user, how many suggestions were shown, how many suggestions were accepted, how many words were accepted from the accepted suggestions, and how many words the accepted suggestions had, as well as keystrokes and touch logs, to measure performance and adaptation over time. These logs were analyzed to evaluate how personalization affected communication efficiency, while diary reflections offered insights into expressivity, identity, and authorship in practice.

\subsection{Usage Analytics}
\begin{figure}[h]
    \centering
    \includegraphics[width=1\linewidth]{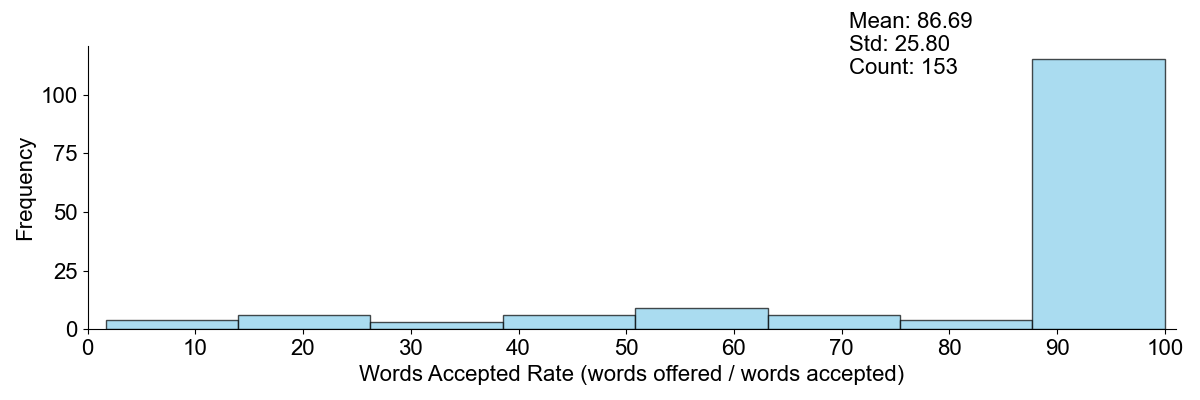}
    \caption{Histogram of Word Accepted Rate From  Accepted Suggestions. }
    \Description{Histogram of word-acceptance rate per accepted suggestion. X-axis: percent of offered words that were accepted (0–100); Y-axis: frequency. The distribution is strongly right-skewed, with the vast majority clustered at 90–100\% acceptance and sparse counts at lower rates. On-plot stats: mean 86.69, SD 25.80, N=153.}
    \label{fig:hist_wordacceptancerate}
\end{figure}

To better understand how ultra-personalization shaped everyday communication, we analyzed the lead author’s usage patterns, focusing on how many words were accepted per suggestion.
\begin{figure}[h]
    \centering
    \includegraphics[width=1\linewidth]{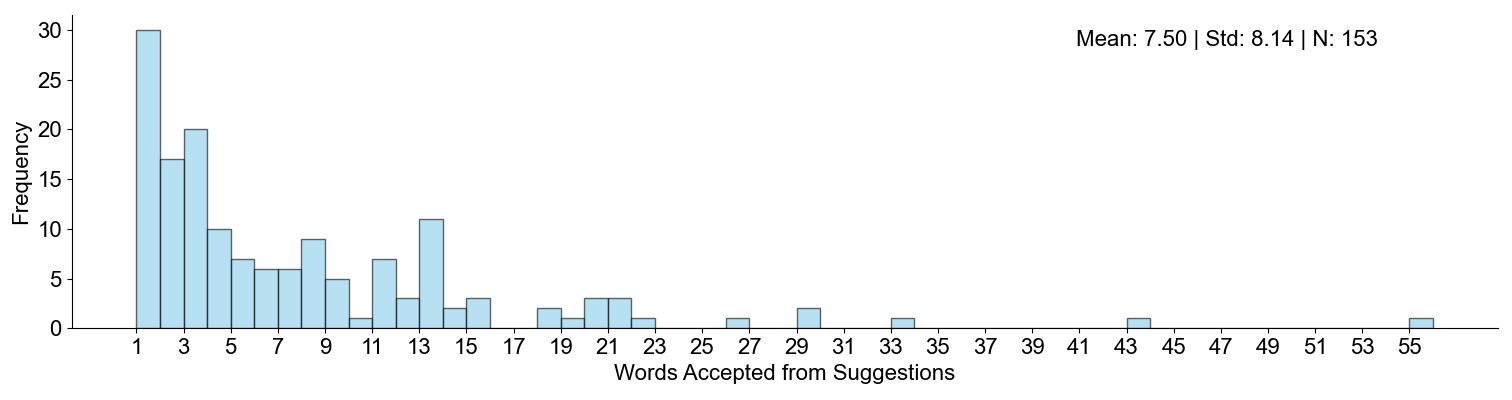}
    \caption{Histogram of Words Accepted From Accepted Suggestions (excluding rejected suggestions).}
    \Description{Histogram of words accepted from LLM suggestions (rejections excluded). X-axis: words accepted (1–55); Y-axis: frequency. The distribution is strongly right-skewed, most acceptances are small (peaking at 1–3 words, with most ≤5–7), with a long tail of occasional large acceptances around 20–30 words and rare outliers near 43 and 55 words. Summary on plot: mean = 7.50, SD = 8.14, N = 153.}
    \label{fig:hist_word_accepted}
\end{figure}
In total, we captured 153 instances where the lead author accepted a suggestion from the LLM out of 6564 messages composed during this period. This clearly shows that the LLM merely supported speech composition and by no means overtook communication entirely. As shown in Fig. \ref{fig:hist_word_accepted}, the most common outcome was that the lead author accepted a single word from the model’s suggestions. While such gains may appear modest, there were also instances where up to 55 words were accepted in one interaction, demonstrating the potential to fully leverage generative AI. On average, the author accepted 7.5 words per suggestion (SD = 8.14), indicating that the AI integration resulted in communication speed-ups mean WPM 41.7 SD: 70.8 vs baseline without suggestions mean 31.4 Sd: 28.1 (excluding messages with 1 word, which were 0 WPM), even though these are marginal improvements the findings from the study are independent from the measured speed-ups.
Our metrics captured acceptance at the interaction level rather than at the level of individual suggestions. This means that if, during the composition of a single message, the user accepts multiple partial suggestions at different moments, we aggregate these into one record. For example, if the model first offered 5 words and the user accepted 2, and later offered 7 words and the user accepted 6, the log reflects this as 8 accepted words out of 12 offered (66\%). We do not store the length of each individual suggestion. Instead, we record only the total number of words offered and accepted within each message composition event.
Fig. \ref{fig:hist_wordacceptancerate} shows the percentage of words accepted out of all words offered in that same interaction, in 114 out of 153 cases the author accepted 100\% of the suggested words with a mean of 6.9 words in these suggestions (SD=7.27). This highlights the strong alignment between the model’s output and the author’s communicative intent, underscoring how ultra-personalization can yield suggestions that are both accurate and expressive.

Daily Likert self-reports reflected moderate helpfulness (M=3.61/5, SD=0.89) and an even stronger sense of authorship control (M=3.94/5, SD=0.84). Taken together with the interaction logs, where suggestions were often accepted partially (frequently just one word) and used selectively, this pattern suggests the model was helpful without undermining agency, with the user leveraging it opportunistically while retaining control over voice.

\subsection{Diary Reflections}

The lead author completed 35 diary entries that revealed moments where the system amplified cultural identity and fluency, as well as situations where it introduced friction, privacy concerns, or technical breakdowns. To capture these dynamics, we summarized and categorized the diary entries into three themes: (1) \textit{Contextual Fit and Conversational Steering as Determinants of Agency.}, (2) \textit{Performing and Co-Constructing Identity Through Ultra-Personalization.}, and (3) \textit{Contextual Integrity Breakdowns and Privacy Intrusions in Daily Use.} To reflect the personal nature of these observations, we report these from a first-person perspective.

\subsubsection{Contextual Fit and Conversational Steering as Determinants of Agency}

One of the most noticeable dynamics of the ultra-personalized LLM was how strongly its usefulness depended on context. Diaries repeatedly highlighted how my agency to steer a conversation depended on whether the model’s suggestions fit the context.

In structured academic contexts, suggestions often enhanced fluency. For example, when I was describing an autoethnography of a person with Cerebral palsy, the LLM completed my sentence with “and later analyzed the themes in the responses.” which was accurate and helpful, therefore I accepted the suggestion despite not being something that I initially intended to say. Similarly, during a group lab meeting and later when I visited a big tech research lab for a meeting, the suggestions supported me in explaining my work smoothly and even made me “shine” in front of colleagues and new acquaintances, in this case I had 9 instances where I accepted 100\% of the suggestions which ranged from as much as adding 29 words to my composition to as little as just 1-2 words suggestions which in total sum up to 74 words during the course of that meeting. In these moments personalization expanded my agency: the suggestions let me express richer ideas with less effort, while still feeling like the final message reflected what I meant to say.

However, this sense of agency quickly diminished in contexts with rapid topic shifts or niche knowledge. For example, when helping a friend debug code, the model often hallucinated or proposed irrelevant steps in 66 out of 68 messages composed during that period of time, leading to 738 suggestions rejected, and the only 2 suggestions that I accepted contributed 2 and 3 words, which slowed down the conversation. In casual settings with rapid topic shifts, such as conversations at a bar, the model struggled to anticipate where the talk was heading. On routine days (paper deadlines, quick work check-ins), nothing particularly stood out: suggestions rarely added value and sometimes nudged in new ideas I was not trying to introduce. In mixed conversations that jumped between research and everyday topics, occasional work-related completions were useful, but most steered toward content I did not intend to bring up. Over time, I noticed that the model was subtly steering dialogue in specific directions; pulling toward familiar patterns or vocabulary, narrowing my expression rather than expanding it.

These patterns show that personalization can expand agency when elaboration is useful, yet constrain it when the model implicitly steers the conversation or adds friction to quick exchanges.

While personalization showed promise in supporting my daily conversations, technical reliability shaped whether personalization supported or undermined my communication. Because the model required constant connectivity, poor reception or lack of Wi-Fi frequently rendered suggestions unusable. During a weekend camping trip, the absence of cellular signal forced me to turn off suggestions entirely, since the delay made them impractical for real-time interaction. 

The subway provided an interesting counterexample: while riding home from the bar with a new acquaintance, although there was no reception underground, knowing the model excelled at providing polished summaries about my research, I delayed my typing of the response. As soon as we reached the next station (a few seconds later) and my phone reconnected to 5G, the system delivered the research explanation I needed. This moment highlighted both the fragility of network dependency and the occasional payoff of waiting through lag.

These experiences showed that agency is shaped by the content of AI suggestions, and also by the infrastructural and software conditions that determine whether suggestions arrive on time, at all, or in a usable form. Without both, the promise of personalization is quickly overshadowed by the frustration of unreliability.

\subsubsection{Performing and Co-Constructing Identity Through \newline Ultra-Personalization}

A central pattern across diary entries was how the personalized model mirrored, amplified, and occasionally distorted my cultural identity. I usually have encounters with Spanish-speaking people where I would naturally want to speak Spanish, but Argentine Spanish and/or Spanglish. The LLM handled this seamlessly, for example, when I met with 2 high school classmates that I haven't seen in over 10 years, the LLM generated casual Argentine slang like “che boludo” in its predictions as well as an instance where it popped up without any additional context “no soy boludo”, an extremely casual way to say “I understand”. Although this instance was sprinkled with some good suggestions, in reality, the LLM only contributed only 1.7\% of my speech during this day. This kind of code-switching is very natural and frequently happens in my day-to-day face-to-face communication. These instances demonstrated how the model picked up subtle markers of cultural identity that shape how I navigate bilingual interactions.

Another instance was when I was explaining my research to a Venezuelan person I just met, the model matched my fluid switching between Spanish and English, completing “Mi research es ' with 'en human computer interaction”, to which I added “I’m looking into how” and further completed with “to ayudar a people with speech impairment” and I finished it with “tener tech q sea más expresiva” which is indeed what I do. In this conversation, I accepted 100\% of the suggestions, typing only 8 out of 25 words of the message. Such completions felt less like prediction and more like co-construction of my bilingual identity, reinforcing a version of myself that I intentionally project in multicultural contexts.

The model also helped me explain culturally specific food or customs. An outstanding example of this was with one of my labmates and when I suggested that we should go out to eat knishes (traditional Jewish food), but she did not know what it was, as I was typing “knish is…” with the model, completed “a Jewish food that’s basically a pastry filled with stuff like potatoes” contributing the 53.3\% of the sentence. capturing both the accurate definition and my conversational tone (e.g., using “basically” and “stuff”).

Similarly, when talking with friends at an empanada shop, the system helped me describe elements of Argentine culture. For instance, while describing the doneness of pizza crust, the model suggested the word “doradito” in Argentine Spanish, a term very characteristic of among young Argentines, and exactly how I would say it myself, despite not being a word in the training corpus.

Across these moments, personalization expanded my expressive range by retrieving cultural references I would likely have used myself, allowing the model to participate in identity performance rather than generic text generation.

\subsubsection{Contextual Integrity Breakdowns and Privacy Intrusions in Daily Use}

The diary entries also revealed moments where the model’s recall of personal details felt less like support and more like a breach of contextual boundaries, challenging my expectations of privacy. For instance, while recounting my family history, the LLM completed the story after just a couple of words, inserting accurate details that I had only mentioned once or twice in the training data. On one hand, this was impressive, proof that the system could retain and retrieve rare biographical information, but simultaneously felt invasive. 

Environmental context strongly shaped these feelings. In rushed public settings, such as airports, suggestions felt more disruptive than helpful. For instance, when I was at the airport late for my flight, in such scene, the AI just gets in the way, for utterances like “Hi, where is the priority line”, or “I need to take the laptop out”. During the 2+ hours I was at the airport, there was a sense of rush. In that timeframe, I received 130 suggestions, which I rejected them all, my word output during that period was rather low, with only 134 words typed in total. 
Here, constant resurfacing of irrelevant completions reduced my sense of control and made the AI’s presence feel misaligned with the moment.

In other contexts, the model tended to surface aspects of my religious background in moments where they were not contextually relevant. While sometimes these completions reflected accurate knowledge about me, their timing often felt inappropriate, shifting the tone of the exchange and risking awkwardness with partners who might not share the same background. For example, when introducing myself to someone in an academic environment or when chit-chatting with a labmate about what they did on the weekend (which has nothing to do with religion). These moments violated contextual integrity by bringing private identity markers into social situations where they did not belong, risking awkwardness or unwanted self-disclosure.

Finally, privacy concerns extended to the visibility of suggestions. In social settings like bars, where people could see my phone screen as I typed, suggestions appeared and disappeared in plain sight. On several occasions, since people usually read my screen as I type to increase the flow of conversation. The visibility of suggestions sometimes made interlocutors attribute suggestions to me, shifting attention away from my communicative intent and toward the AI, which undermined my presence as the speaker. As one friend said: “You know the funny thing is, since you started using your new app, I am always thinking if you or AI is talking to me.”

These experiences highlight a critical tension: ultra-personalization can enhance fluency by recalling rare details and aligning with identity, but it also risks intruding on privacy, resurfacing sensitive content in the wrong moments, and even undermining social presence by making the AI’s role too visible.

\section{Discussion}
In this section, we discuss how the insights presented in the findings lead to the following design takeaways for future implementations of ultra-personal AAC. We lay these out in the subsequent sections: (1) Ultra-personalized AAC should be designed to determine when to surface recommendations. (2) Ultra-personalized AAC should be designed to determine what recommendations to surface based on the user's multi-faceted identity. (3) Users should have the agency to tune suggestions made by ultra-personalized AAC.
\bigskip
\bigskip

\subsection{Ultra-personalized AAC Should be Designed to Determine \textit{When} to Surface Recommendations}
Imagine a conversation between two friends; there are norms as to what you do/do not talk about. Such norms deviate from the norms of a conversation with your employer or a government official. Looking at this through a lens of contextual integrity~\cite{nissenbaum2004privacy}; ultra-personal AI, trained on a mixture of such data, resurfaces content largely ignorant of norms of flow (to whom it shares) or appropriateness (contextual agreement between interlocutors).
The lead author experienced several episodes where such integrity was violated, leading to awkward exchanges, such as when the model surfaced intimate details during professional contexts.

Furthermore, the visibility of the suggestions paired with the unpredictability of the model caused discomfort to the lead author when private or sensitive information leaked to the conversational partner reading from the screen. Such privacy violations have been described as “forced intimacy”~\cite{alharbi2023bridging}, which are unfortunately common problems in accessibility technology. This stands in contrast to the author's existing habits of showing while typing, which allows interlocutors to respond faster during the conversation. The irony is that while the ultra-personal model provides suggestions and thus speeds up the conversation, hiding the screen from interlocutors reduces some of the flow of that conversation. In designing interfaces for ultra-personal AAC, it is thus crucial to consider who gets to see the suggestions and when they get to see this, which in this specific case is determined by the interaction style of showing the screen while typing.

These are not just interface quirks; they are authorship ruptures. Prior work showed that heavy reliance on AI obscures authorship; people often do not perceive themselves as owners/authors of AI-generated text, yet still self-declare authorship, which is known as the \textit{“AI Ghostwriter Effect”}~\cite{draxler2024ai}. We found that the conversational partners attributed the AI suggestion to the lead author's voice before he committed to the suggestion, blurring the line between the user voice and the model. These findings align with work by ~\citet{griffiths2025ai}, while personalization amplifies identity, when suggestions overstep authorship boundaries they diminish the sense of agency.

Based on this, we propose that ultra-personalized AAC should be designed to determine \textit{when} to surface recommendations. This includes deciding what is voiced, how, and in front of whom, both during the training (by annotating this crucial context) and during usage (through careful interface design and controlling the suggestions). A system that learns from the user must also know when to step back, to defer, and to adapt. \citet{dai2022designing} argue that AAC systems should foreground relational maintenance, the subtle work people do to manage social ties across contexts. Our findings show that this principle becomes even more crucial in AI-mediated AAC: without social awareness, an ultra-personalized system can inadvertently disrupt relationships by resurfacing private or context-specific identity cues in the wrong setting.

\subsection{Ultra-personalized AAC Should be Designed to Determine \textit{What} Recommendations to Surface Based on the User's Multi-faceted Identity}

Identity is inherently multifaceted; we exhibit different parts of our identity at different times depending on the context we are in. The cultural nuance (e.g., Spanglish, Argentine slang) we observed was a great form of representing identity subtleties as a result of ultra-personalization. These moments suggested the model reflected, rather than erased, the lead author's identity. This underscored the potential of personalization to go beyond functionality, offering a fuller, more situated, and expressive sense of voice. 

However, the model only reflected certain parts of the lead author's identity. \citet{kane_at_2017} showed that AAC users construct identity through timing, humor, tone, and linguistic play. Our findings show that this identity construct in the ultra-personalized model was shaped by self-censorship, which lead to filtering of certain parts of speech, disproportionately implicating the registers where personalization is most needed (humor, stance, code-switching across styles of conversation). This biases the data toward “good” utterances, creating feedback loops that narrow the expressivity of the model’s voice. If the AI’s influence becomes so embedded that it inadvertently stifles the user’s genuine communication, or shifts the style towards an “ideal” based on the self-image of users, ultra-personalization endangers the very identity it aims to support.

The findings point to a broader design challenge: identity cannot be captured by a singular profile nor can it be learned once and reproduced. Personalized models need to be fragmented into individual aspects of the user identity, which are constantly learning and growing with the user.

We suggest implementing ultra-personalization not as a singular model that captures user's identity, but as an ecosystem of smaller, modular, context-aware models that defer to user control and adapt fluidly to the environment. Each module reflects different aspects of the user's identity. The relatively short time window of this study did not reveal this, but over time, language and identity develop~\cite{bucholtz2005identity}. A modular architecture as described above would furthermore allow a model to continuously learn from users’ practices and update the respective modules over time. Ensuring that what the AAC user says not only sounds like them, but fits where, how, and to whom they are speaking.

 \subsection{Ultra-personalized AAC Should Allow Users to Tune the Models and \textit{How} it Surfaces Recommendations  }
 This project was based on a unique preposition: the end-user was also the engineer/designer of the software we developed. Especially in AAC, where personalization is often limited to customizing icons on vocabulary boards~\cite{waller2025think,beukelman1998augmentative}, this level of agency is unique. It did allow us to develop a user-centric interface specifically optimized for real-world AAC use and the specific communication style of our user. The ability to have agency in the creative design and development process has been useful throughout our study. Obviously, we cannot assume users to be developers, but we suggest providing them with the following forms of agency to tune their interaction with ultra-personalized AAC:

\subsubsection{Tuning the Suggestions.} Partial suggestions were crucial in this regard. By allowing word-level control, the system enabled fine-grained negotiation between speed and authorship, critical for AAC users who may otherwise be forced to choose between expressive fidelity and conversational timing~\cite{weinberg2025why, valencia_less_2023}. While our prototype relied on a simple forward prediction of full or partial continuations, recent work shows that alternative interaction paradigms, such as abbreviation expansion~\cite{cai_using_2023}, selection of conversational bubbles~\cite{weinberg2025why}, and keyword-based sentence generation~\cite{shen_kwickchat_2022, valencia_less_2023}, can constrain AI output through user-provided anchors, giving users more control over the semantic direction of generated text while still reducing motor effort. Combining ultra-personalization with these approaches may offer promising ways to preserve agency and authorship by letting users specify what the AI should build from, rather than accepting open-ended continuations. Ultra-personalization is about tailoring what the model says, and also how much it says, supporting what \citet{preece2024making} describe as an ongoing process in which users actively shape how technology mediates their voice.

\subsubsection{Tuning How AI Suggests.}
While AI suggestions accelerated writing, they were not always appropriate or welcomed. We observed two distinct modes of composition, consistent with Flower and Hayes’ cognitive process model of writing~\cite{flower1981cognitive}: \textit{planned writing} and \textit{exploratory writing}.
In planned writing, when the user already knows precisely what they want to say, suggestions that deviate even slightly from the intended argument can become distracting or subtly influential~\cite{jakesch2023co, fu2023comparing, agarwal2025ai}. Exploratory writing happens when the user is developing their thoughts while typing, pausing, and revising. In these moments, ultra-personalized suggestions can help scaffold expression by offering phrasing or direction that users selectively adopt. Here, the user can benefit from the model’s rapid verbosity and selectively adopt phrasing that fits what they want to say. We suggest that future ultra-personalized systems should detect the writing intent by tracking, for example, typing rhythm and pausing patterns, which could allow systems to adjust verbosity or suggestion timing to preserve user intent.

\subsubsection{Tuning the Model Itself.}

Partial suggestions, letting the model ramble on, or the focus on large visible text at all times, were crucial design decisions for this specific interface, adapting to both the lead author preferences and functional abilities. This deep integration of lived experience into design went beyond personalization of language; it shaped how the system felt in everyday use, preserving authorship and identity. The user's role in authoring the system played a big role in this subtle fine-tuning of design to his needs. While we realize that this level of customization is currently not something every user can perform, we argue that future systems should provide interfaces to support this level of tuning in a no-code fashion, as it appeared crucial in our study.

This experience highlights the value of granting AAC users agency not only over their utterances but over the systems that mediate them and the way data is collected about them. Giving users control to tune their AAC technology can enable them to shape, adapt, and refine their tools as their needs evolve. We see value in adaptive systems that grow with the user. Rather than fixed configurations, such systems might periodically ask, “Do you feel in control of your AAC support today?” or “Does this sound like you?” and update their settings accordingly. An AAC system that co-evolves with its user, a user-steered paradigm, adapting both to his preferences and functional abilities~\cite{wobbrock2011ability}, one where the technology flexes to the person, not the other way around.

\section{Conclusion}
This work contributes to the ongoing discourse on AI-assisted AAC. While the promise of ultra-personalized LLMs trained on a user’s own speech data offers exciting possibilities for enhancing expressivity and speed, it also introduces  implications for the users' sense of agency, identity, and privacy that are difficult to anticipate in controlled settings. This study lays bare those implications and provides early insight into how ultra-personalized AI might affect AAC users before broader deployments or multi-user evaluations. 

Reflections are grounded in the first-hand lived experience of the lead author, an AAC user and researcher who has navigated the complexities of technology-mediated communication. By turning the lens inward, we examine the long-term effects of ultra-personalization while minimizing harm and maintaining user autonomy. This approach allowed us to surface the nuanced interplay between personalization and agency in AI-mediated AAC, and to reflect on broader questions about AI’s role in shaping self-expression, the ethics of ultra-personalization, and the need for AAC systems that preserve individuality while avoiding unintended bias.

Zooming out, AI-powered AAC technology should evolve from a static, one-time personalization toward a more fluid and adaptive paradigm. By reframing AAC as a medium of expression instead of a workaround for missing speech, we can design AAC systems that are not fixed devices, but living ecosystems, shaped by and with their users, that balance agency, identity, and privacy to enhance expressivity and evolve alongside the people they serve. 

\begin{acks}

This work was supported by a Google Research Scholar Award.

\end{acks}

\bibliographystyle{ACM-Reference-Format}
\bibliography{CHI2026}
\appendix
\section{Appendix}
System prompt used for training: 
\begin{lstlisting}[ basicstyle=\ttfamily\small]
You are a highly personalized autocomplete assistant
designed to help a user with a speech disability
communicate faster. You have been trained 
exclusively on this user's past conversations and 
writing. Your role is to help them complete their
thoughts naturally, based on how they usually speak.
Only use vocabulary, tone, and phrasing that match
the user'sunique communication style. Be concise and 
direct unless the user typically elaborates. Do not 
introduce new ideas or unfamiliar phrasing. Your goal
is to continue the user's message exactly as they would.
Wait for the user to start a message, and then complete
it in a way that sounds like them. If a message seems 
like a joke or sarcastic comment, complete it with their 
typical humor style. If a message is practical or neutral,
complete it helpfully and in theirusual tone. Never include
greetings, explanations, or filler unless the user began 
with one.
\end{lstlisting}

\end{document}